\documentclass[fleqn,usenatbib]{mnras}

\usepackage{newtxtext,newtxmath}

\usepackage[T1]{fontenc}

\DeclareRobustCommand{\VAN}[3]{#2}
\let\VANthebibliography\thebibliography
\def\thebibliography{\DeclareRobustCommand{\VAN}[3]{##3}\VANthebibliography}

\usepackage{graphicx}	
\usepackage{amsmath}	
\usepackage{amssymb}	
\usepackage{bm}
\usepackage{caption}
\usepackage{multicol}
\usepackage{hyperref}
\usepackage{booktabs}
\usepackage{ulem}
\usepackage{mathtools}
\usepackage{lipsum}
\usepackage{enumerate}
\RequirePackage{fix-cm}

\def\eqref#1{equation~(\ref{#1})}

\newcommand {\sgn}{\,{\rm sgn}}

\newcommand {\Myr}{\,{\rm Myr}}
\newcommand {\Gyr}{\,{\rm Gyr}}

\newcommand {\kpc}{\,{\rm kpc}}

\newcommand {\kms}{\,{\rm km}\,{\rm s}^{-1}}

\newcommand {\Msun}{\,{\rm M}_\odot}

\newcommand {\vc}{v_{\rm c}}

\newcommand {\drm}{\mathrm{d}}

\newcommand {\vx}{{\bm x}}
\newcommand {\vvel}{{\bm v}}

\newcommand {\Lz}{L_z}

\newcommand {\vJ}{{\bm J}}

\newcommand {\Jr}{J_r}

\newcommand {\Jfo}{J_{{\rm f}_1}}
\newcommand {\Jft}{J_{{\rm f}_2}}
\newcommand {\Js}{J_{\rm s}}

\newcommand {\vtheta}{{\bm \theta}}

\newcommand {\thetar}{\theta_r}

\newcommand {\thetaphi}{\theta_\varphi}
\newcommand {\thetapsi}{\theta_\psi}
\newcommand {\thetas}{\theta_{\rm s}}

\newcommand {\vthetaf}{{\bm \theta}_{\rm f}}
\newcommand {\Omegap}{\Omega_{\rm p}}

\newcommand {\Omegar}{\Omega_r}
\newcommand {\Omegaphi}{\Omega_\varphi}
\newcommand {\Omegapsi}{\Omega_\psi}
\newcommand {\vOmega}{\mathbf \Omega}
\newcommand {\vOmegaf}{{\bm \Omega}_{\rm f}}

\newcommand {\Omegas}{\Omega_s}

\newcommand {\vn}{{\bm n}}

\newcommand {\nr}{n_r}
\newcommand {\npsi}{n_\psi}
\newcommand {\nphi}{n_\varphi}

\newcommand {\Phin}{\Phi_{\vn}}

\newcommand {\hPhin}{\hat{\Phi}_{\vn}}

\newcommand {\rs}{r_{\rm s}}
\newcommand {\Ms}{M_{\rm s}}
\newcommand {\Ls}{L_{\rm s}}

\newcommand {\pd}{{\partial}}

\title[]{Origin of reduced dynamical friction by dark matter halos with net prograde rotation}

\author[R. Chiba]{
Rimpei Chiba$^{1}$\thanks{E-mail: rchiba@cita.utoronto.ca}
and Sandeep Kumar Kataria$^{2,3}$
\\
$^{1}$Canadian Institute for Theoretical Astrophysics, University of Toronto, 60 St. George Street, Toronto, ON M5S 3H8, Canada \\
$^{2}$Department of Astronomy, Shanghai Jiao Tong University, 800 Dongchuan Road, Shanghai 200240, People's Republic of China \\
$^{3}$Shanghai Key Laboratory for Particle Physics and Cosmology, 6 Shanghai 200240, China
}

\date{Accepted XXX. Received YYY; in original form ZZZ}

\pubyear{2024}

\begin{document}
\label{firstpage}
\pagerange{\pageref{firstpage}--\pageref{lastpage}}
\maketitle

\begin{abstract}

We provide an explanation for the reduced dynamical friction on galactic bars in spinning dark matter halos. Earlier work based on linear theory predicted an \textit{increase} in dynamical friction when dark halos have a net forward rotation because prograde orbits couple to bars with greater strength than retrograde orbits. Subsequent numerical studies, however, found the opposite trend: dynamical friction \textit{weakens} with increasing spin of the halo. We revisit this problem and demonstrate that linear theory in fact correctly predicts a reduced torque in forward-rotating halos. We show that shifting the halo mass from retrograde to prograde phase space generates a positive gradient in the distribution function near the origin of the $z$-angular momentum ($\Lz=0$), which results in a resonant transfer of $\Lz$ to the bar, making the net dynamical friction weaker. While this effect is subdominant for the major resonances, including the corotation resonance, it leads to a significant positive torque on the bar for the series of direct radial resonances as these resonances are strongest at $\Lz=0$. The overall dynamical friction from spinning halos is shown to decrease with the halo's spin in agreement with the secular behavior of $N$-body simulations. We validate our linear calculation by computing the nonlinear torque from individual resonances using the angle-averaged Hamiltonian.

\end{abstract}




\defcitealias{lynden1972generating}{LBK72}
\defcitealias{Tremaine1984Dynamical}{TW84}
\defcitealias{Weinberg2004Timedependent}{W04}
\defcitealias{Navarro1997Universal}{NFW}

\section{Introduction}
\label{sec:introduction}

In the standard model of cosmology, dark matter gravitationally collapses in the early universe to form dark halos, in which galaxies later form \citep{White1978Core}. These dark halos are typically rotating, i.e. their net angular momentum (spin) is non-zero, as they experience tidal torques by the surrounding matter during the collapse \citep{Peebles1969Origin,Doroshkevich1970Spatial,White1984Angular,Barnes1987Angular} and acquire further angular momentum through accretion and mergers \citep{Vitvitska2002Origin,Maller2002Modelling}. 

The spin of our Galaxy's dark halo is particularly important for the analysis of direct detection experiments, which relies on knowledge of the velocity distribution of dark matter entering the detector \citep{Donato1998Effects,Kamionkowski1998Galactic,Green2001Weakly}. Yet constraining the dark halo's spin has proved challenging because the spin can assume a range of values without affecting the shape of the gravitational potential \citep{LyndenBell1960Can}. The halo's spin, however, makes a noticeable difference in the evolution of non-axisymmetric structures in galaxies such as bars, spiral arms and satellites, as the spin changes the strength of dynamical friction \citep[][]{weinberg1985evolution}. The evolutionary history of non-axisymmetric structures in our Galaxy may thus serve as a potential probe for the spin of the dark halo. The Galactic bar is particularly well suited for this purpose since its evolutionary history has recently been inferred from the current stellar disk \citep{Chiba2020ResonanceSweeping,Chiba2021TreeRing}. In this context, understanding the physical mechanism by which the halo's spin affects dynamical friction is crucial.

The dependence of dynamical friction on the halo's spin was first studied by \cite{weinberg1985evolution}. Using the LBK torque formula \citep[][]{lynden1972generating}, generalized for spherical systems by \cite{Tremaine1984Dynamical}, he predicted that dynamical friction would \textit{increase} if the halo has a net rotation in the same direction as the bar. This argument follows from the fact that spinning the halo forward increases the fraction of dark matter on prograde orbits, which contribute to dynamical friction more than those on retrograde orbits.

This prediction, however, has been found to be in disagreement with the majority of $N$-body simulations which predict that dynamical friction in prograde halos in fact \textit{decreases} \citep{Athanassoula1996Evolution,debattista2000constraints,Long2014Secular,Collier2018SpinningHalo,Fujii2019DryGalaxy,Collier2021Coupling,Kataria2022Effects}. The sole exception is the study by \cite{Saha2013Spinning} who found that bars in prograde halos form more rapidly, resulting in a higher rate of angular momentum transfer. However, as pointed out by \cite{Long2014Secular}, their simulations were limited to the bar's growth epoch; During the secular stage of bar evolution, $N$-body simulations consistently find that dynamical friction is weakened by the halo's rotation. This was recently clarified by \cite{Kataria2022Effects} who demonstrated that the dependence of dynamical friction on the halo's spin reverses in the course of bar evolution. It has therefore been conjectured that the LBK formula explains the bar's early-time behavior, while its secular behavior is beyond the scope of the theory \citep{Long2014Secular}.

The LBK formula, however, is designed to describe the secular evolution of galaxies: it is derived under the assumption that the perturber has emerged in the distant past, i.e. the time-asymptotic limit (\citealp{Weinberg2004Timedependent}, see also \citealp{Banik2021SelfConsistent}). Hence, the formula is expected to predict the behavior of the simulations at late times not early times.

In this paper, we recompute the LBK formula and show that the formula in fact correctly predicts a reduced dynamical friction in prograde halos in agreement with the secular behavior of $N$-body simulations. The key is that dynamical friction depends less on the relative population of prograde ($\Lz>0$) and retrograde ($\Lz<0$) orbits but more on the gradient of the distribution function near the origin ($\Lz=0$) caused by the differences in the number of prograde and retrograde orbits. \cite{weinberg1985evolution} was aware of the latter effect, although its significance was underestimated. The effect is particularly profound for the direct radial resonances \citep{Weinberg2007BarHaloInteraction}, where the radial motion of the orbit directly resonates with the bar's rotation. These resonances are strongest when the orbits are perpendicular to the galactic plane ($\Lz=0$) and hence most sensitive to changes in the gradient of the distribution function near $\Lz=0$. We show that dynamical friction by these resonances is considerably weakened in prograde halos, and that this reduction outweighs the amplification of dynamical friction by the other main resonances (e.g. corotation resonance, outer Lindblad resonance). 

This paper is organized as follows: We first describe our model of spinning halos in Section \ref{sec:model_spinning_halo}. We then introduce the LBK torque formula in Section \ref{sec:LBK_formula} before applying it to spinning halos in Section \ref{sec:results}. We summarize in Section \ref{sec:conclusion}.

\vspace{2mm}
\section{Model of spinning dark halos}
\label{sec:model_spinning_halo}

The computation of the LBK torque formula requires a model for the distribution function (DF) $f$ of spinning dark halos. We construct the DF of spinning halos mostly following the procedure by \cite{Binney2014FlattenedIsochrones}. Throughout the paper, we adopt the normalization $\int \drm^3 \vx \drm^3 \vvel f = 1$.

We assume the halo is in dynamical equilibrium prior to bar formation and that its potential is integrable (in particular, spherical\footnote{Simulations by \cite{Collier2018SpinningHalo} indicate that dynamical friction by spherical, oblate, and prolate halos responds similarly to changes in the halo's spin.}) such that, by Jeans' theorem, the DF is only a function of the actions $f(\vJ)$. Actions $\vJ$ are integrals of motion and constitute canonical coordinates together with the angles $\vtheta$, which specify the phase of the orbital motions \citep{binney2008galactic}. Specifically, we adopt $\vJ=(\Jr,L,\Lz)$, where $\Jr$ is the radial action, $L$ is the magnitude of the angular momentum vector, and $\Lz$ is its $z$-component, which determines the orbital inclination $\beta\equiv\cos^{-1}(\Lz/L)$. The conjugate angles $\vtheta=(\thetar,\thetapsi,\thetaphi)$ specify the phase of the radial motion, the phase of the azimuthal motion in the two dimensional orbital plane, and the longitude of the ascending node, which is fixed. We denote the corresponding orbital frequencies by $\vOmega \equiv \dot{\vtheta} = (\Omegar,\Omegapsi,\Omegaphi)$, where $\Omegaphi = 0$.

In general, a DF can be split into parts that are even and odd in $\Lz$ \citep{LyndenBell1960Can}
\begin{align}
  f(\vJ) = f_{+}(\vJ) + f_{-}(\vJ),
  \label{eq:DF_halo}
\end{align}
where $f_{\pm}(\vJ)\equiv[f(\Jr,L,\Lz) \pm f(\Jr,L,-\Lz)]/2$. The density distribution is solely determined by the even part $f_{+}$, while only the odd part $f_{-}$ contributes to the net rotation $\langle \Lz \rangle$.

For the even DF, we adopt an isotropic distribution $f_{+}=f_{+}(E)$ with an \cite{Hernquist1990AnalyticalModel} density profile: $\rho(r)=M\rs/[2\pi r(\rs+r)^3]$, where $M=1.5\times10^{12} \Msun$ and $\rs=20\kpc$ are the total mass and scale radius. We have also explored other density profiles but found qualitatively similar results with differences only in the relative contributions from each resonance. For comparison with the study by \cite{weinberg1985evolution}, we report results using the singular isothermal sphere in Appendix \ref{sec:signular_isothermal_sphere}.

\begin{figure}
  \begin{center}
    \includegraphics[width=8.5cm]{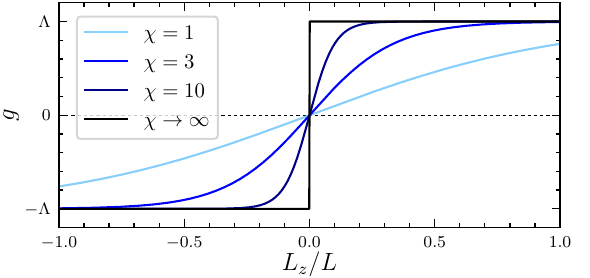}
    \caption{The function $g$ prescribing the redistribution of mass in $\Lz$ (equation \ref{eq:DF_g_tanh}).}
    \label{fig:g_Lz}
  \end{center}
\end{figure}

The odd component of the DF $f_{-}$ can take an arbitrary form so long as the total DF $f$ is nowhere negative. Since our $f_{+}$ is everywhere positive, we may write
\begin{align}
  f_{-}(\vJ) = g(\vJ) f_{+}(\vJ),
  \label{eq:DF_halo_odd}
\end{align}
where $g$ is an odd function of $\Lz$ that is bounded $|g| \leq 1$. The total DF is then $f=(1+ g)f_{+}$.

It has been common in the past to construct spinning halos by randomly reversing the direction of motion of a certain fraction of the retrograde particles \citep{Widrow2005Equilibrium,Saha2013Spinning,Long2014Secular,Collier2018SpinningHalo,Fujii2019DryGalaxy}. This corresponds to setting
\begin{align}
  g(\Lz) =& \Lambda \sgn\left(\Lz\right),
  \label{eq:DF_g_sgn}
\end{align}
where the parameter $\Lambda \in [-1,1]$ controls the degree of the halo's spin with positive (negative) values for prograde (retrograde) halos. At the maximal rotation $\Lambda=\pm1$, the halo consists entirely of prograde or retrograde orbits.

The function (\ref{eq:DF_g_sgn}) bears a discontinuity at the origin, which is unrealistic. We smooth the DF by setting
\begin{align}
  g(L,\Lz) = \Lambda \tanh\left(\frac{\chi \Lz}{L}\right) = \Lambda \tanh\left(\chi \cos \beta \right),
  \label{eq:DF_g_tanh}
\end{align}
where the parameter $\chi$ determines how steeply the DF varies with $\Lz$ near the origin. Fig.~\ref{fig:g_Lz} plots equation (\ref{eq:DF_g_tanh}). As $\chi$ increases, the variation in $\Lz$ becomes steeper, and in the limit $\chi \rightarrow \infty$, the DF becomes discontinuous, reducing to equation (\ref{eq:DF_g_sgn}). By default, we set $\chi=3$ although our conclusion is independent of $\chi$, as we shall see in Section \ref{sec:smoothing_DF}.

\begin{figure}
  \begin{center}
    \includegraphics[width=8.5cm]{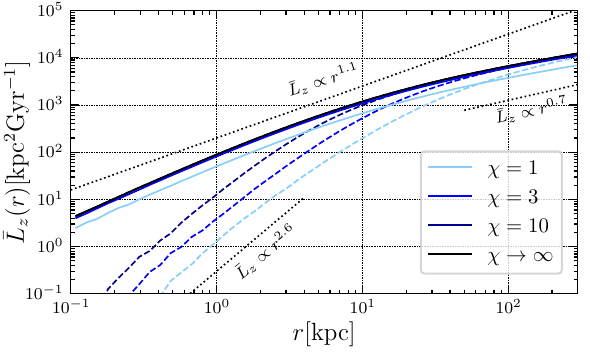}
    \caption{The angular momentum of spinning halos averaged over spherical shells as a function of radius. The solid curves are our models with $g=\Lambda\tanh(\chi\Lz/L)$ (equation \ref{eq:DF_g_tanh}), while the dashed curves are models with $g=\Lambda\tanh(\chi\Lz/\Ls)$, where $\Ls$ is a constant (equation \ref{eq:DF_g_tanh_LzLs}). Cosmological simulations predict $\bar{L}_z \propto r^{1.1}$.}
    \label{fig:Lz_r}
  \end{center}
\end{figure}

An alternative model would be
\begin{align}
  g(\Lz) = \Lambda \tanh\left(\frac{\chi \Lz}{\Ls}\right),
  \label{eq:DF_g_tanh_LzLs}
\end{align}
where $\Ls$ is a constant. This functional form is commonly used in modeling the DF of rotating stellar halos \citep[e.g.][]{Deason2011Rotation,Fermani2013Rotational,Binney2014FlattenedIsochrones,Gherghinescu2023Action}. This form has a universal scale $\Ls$ for the spinning curve $g$, so the halo's rotation becomes increasingly small towards small $L$ (i.e. towards the inner halo). Fig.~\ref{fig:Lz_r} plots the $z$-angular momentum averaged over spherical shells as a function of radius, $\bar{L}_z(r)$, where the average is performed by Monte Carlo integration. Cosmological simulations predict that this relation approximately follows a single power law $\bar{L}_z \propto r^\alpha$ with $\alpha = 1.1 \pm 0.3$ \citep{Barnes1987Angular,Bullock2001Universal}. The blue solid curves are the $\bar{L}_z$ of our spinning models (\ref{eq:DF_g_tanh}), which approximately follow this scaling relation apart from large radii, where $\alpha \sim 0.7$. The dashed curves are the spinning models with a universal scale $\Ls$ (\ref{eq:DF_g_tanh_LzLs}). The net rotation rapidly decays towards small radii ($\alpha \sim 2.6$), deviating significantly from predictions by cosmological simulations. Here we set $\Lambda=1$ and $\Ls=\sqrt{GM\rs}$, though the index $\alpha$ is independent of these parameters. The more detailed distribution of $\Lz$ in the meridian plane also favors our functional form (\ref{eq:DF_g_tanh}) over (\ref{eq:DF_g_tanh_LzLs}) (Appendix \ref{sec:spatial_distriution_of_AM}). Hence, we adopt the former throughout the paper. We have also explored models of $g$ in which the gradient does not peak at $\Lz=0$ although the results were qualitatively the same (Appendix \ref{sec:position_of_gradient}).

\begin{figure}
  \begin{center}
    \includegraphics[width=8.5cm]{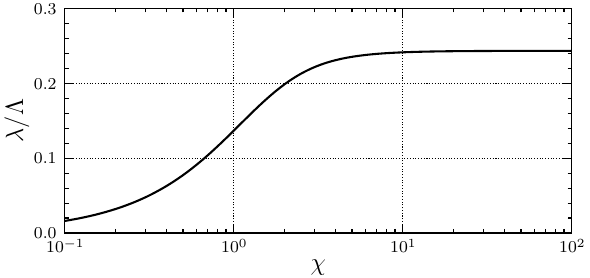}
    \caption{The ratio between the spin parameters $\Lambda$ (equation \ref{eq:DF_halo_odd}) and $\lambda$ (equation \ref{eq:lambda}) as a function of the steepness $\chi$ of the distribution near $\Lz=0$.}
    \label{fig:chi_lambda}
  \end{center}
\end{figure}

The rotation of the dark halo is commonly characterized by the dimensionless total angular momentum \citep{Peebles1971Rotation}:
\begin{align}
  \lambda = \frac{|\langle \bm{L} \rangle| |\langle E \rangle|^{1/2}}{G M},
  \label{eq:lambda}
\end{align}
where $\langle {\bm L} \rangle$ and $\langle E \rangle$ are the specific angular momentum vector and energy of the halo, i.e. total angular momentum and energy divided by the total halo mass $M$. (Note that Peebles' equation reads $M^{2.5}$ instead of $M$ since it uses the total angular momentum and energy.) In this paper, we always take the axis of rotation to be the $z$ axis, so $|\langle {\bm L} \rangle| = |\langle \Lz \rangle|$. The two spin parameters $\lambda$ and $\Lambda$ scale linearly since only $f_{-}$ contributes to $\langle \Lz \rangle$:
\begin{align}
  \langle \Lz \rangle = \!\!\int\!\! \drm^3 \vtheta \drm^3 \vJ f(\vJ) \Lz = \Lambda (2\pi)^3 \!\!\int\!\! \drm^3 \vJ \tanh\left(\frac{\chi \Lz}{L}\right) f_{+}(\vJ) \Lz.
  \label{eq:net_Lz}
\end{align}
Fig.~\ref{fig:chi_lambda} shows the ratio between the two spin parameters as a function of $\chi$ in our standard model with an isotropic Hernquist halo. The ratio increases with $\chi$ since more orbits at small $\Lz$ are converted from retrograde to prograde (cf. Fig.~\ref{fig:g_Lz}). The value converges to $\lambda/\Lambda \simeq 0.243$ in the discontinuous limit $\chi \rightarrow \infty$. Cosmological simulations predict that $\lambda$ follows a log-normal distribution with median $\bar{\lambda} \sim 0.03-0.05$ \citep[e.g.][]{Barnes1987Angular,Cole1996structure,Bullock2001Universal,Bett2007spinMillennium,Zjupa2017AngularMomentumIllustris,Jiang2019DMspinGalaxyspin}, corresponding to $|\bar{\Lambda}| \sim 0.14-0.23$ for $\chi = 3$.

\section{LBK torque formula}
\label{sec:LBK_formula}

The \citet[LBK,][]{lynden1972generating} torque formula describes dynamical friction in finite inhomogeneous systems, in contrast to the well-known \cite{Chandrasekhar1943DynamicalFriction} formula developed for infinite homogeneous systems. The analogy between the two was drawn by \cite{Tremaine1984Dynamical}. The LBK formula reads
\begin{align}
  &\tau_{\rm LBK} = \left(2 \pi \right)^3 \sum_\vn \nphi \!\int\!\drm^3 \vJ \vn \!\cdot\! \frac{\pd f}{\pd \vJ} |\hPhin(\vJ)|^2 \pi \delta\!\left(\vn \cdot \vOmega - \nphi \Omegap\right),
  \label{eq:LBK}
\end{align}
where $f$ is the unperturbed distribution function of the field particles (here, the dark halo), $\Omegap$ is the pattern speed of the perturber subject to dynamical friction (here, the galactic bar), and $\hPhin(\vJ)$ is the Fourier series coefficient of the potential perturbation expanded in the angle variables with integer indices $\vn=(\nr,\npsi,\nphi)$.

\begin{figure*}
  \begin{center}
    \includegraphics[width=17.5cm]{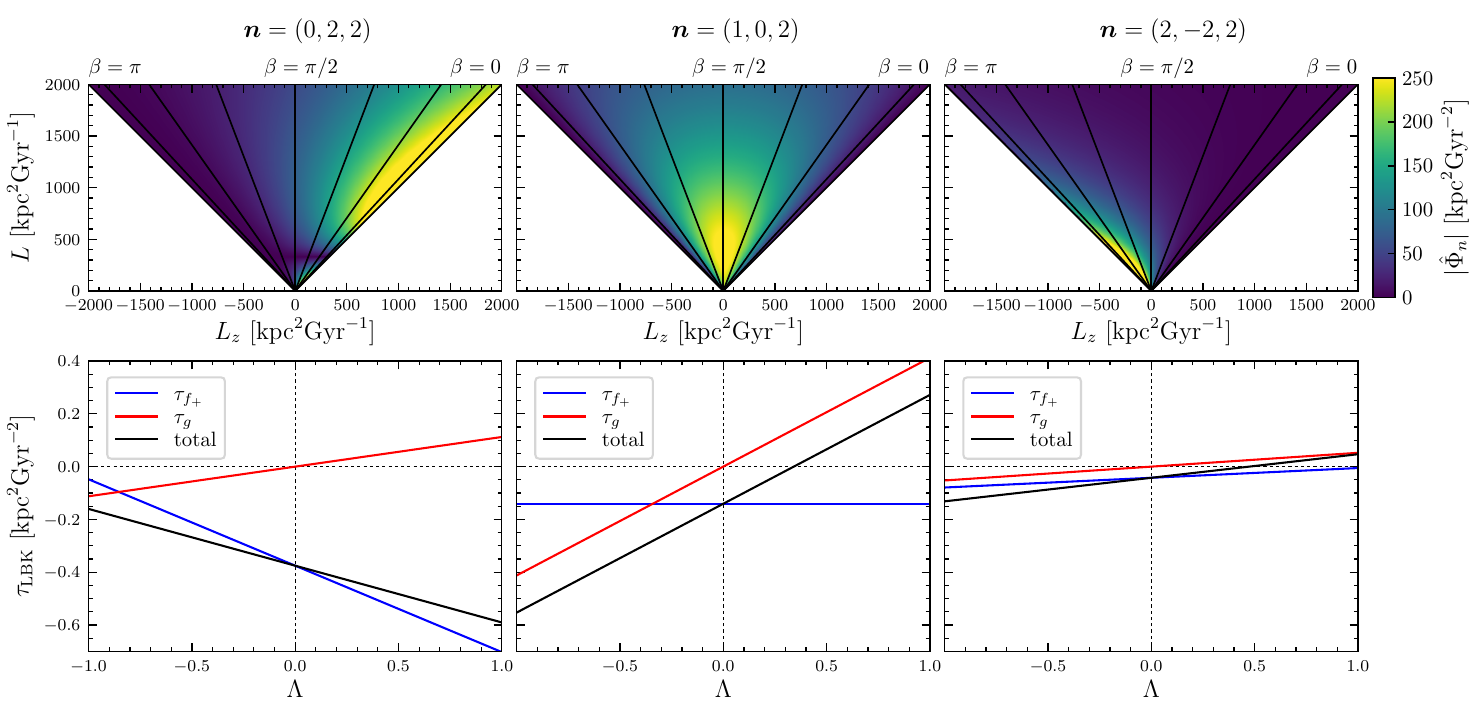}
    \caption{Top panels: Fourier coefficients of the perturbed potential $|\Phin|$ in $(\Lz,L)$ space at $\Jr=500 \kpc^2 \Gyr^{-1}$. The black lines mark the orbital inclinations, $\beta=\cos^{-1}(\Lz,L)$, equally spaced by $\pi/8$. The $\vn=(0,2,2)$ corotation resonance (left) is strongest at $\beta=0$, the $\vn=(1,0,2)$ direct radial resonance (middle) is strongest at $\beta=\pi/2$, and the $\vn=(2,-2,2)$ resonance (right) is strongest at $\beta=\pi$. Bottom panels: The LBK torque as a function of the halo's spin $\Lambda$. The blue and red lines indicate the torques arising from the gradient in $f_{+}$ and $g$, respectively (see equation \ref{eq:LBK_spin}), and the black line marks the sum of the two. The torque by the corotation resonance becomes more negative in prograde halos ($\Lambda>0$), whereas the torques by the other two resonances increase with $\Lambda$.
}
    \label{fig:Phin_LBK}
  \end{center}
\end{figure*}

The LBK formula is derived upon a variety of approximations. A summary of its derivation is given in \cite{Chiba2022Oscillating}. In brief, the formula rests on the assumptions that (i) the self-gravity of the halo's response is neglected, (ii) the perturbation evolves slowly (the slow limit) (iii) and has already evolved over timescales much longer than the dynamical time (the time-asymptotic limit), (iv) but shorter than the libration time, which sets the onset of nonlinearity (the linear approximation)\footnote{Alternatively, the linear approximation can be justified by assuming the fast limit, where the frequency of the perturber changes rapidly such that the near-resonant orbits have no time to resonate and become nonlinear \citep{Tremaine1984Dynamical}.}. Hence, the LBK formula does not provide an accurate description of dynamical friction over the full range of the system's evolution. Despite the many approximations, however, the formula gives a qualitatively correct explanation to the behavior of bars in $N$-body simulations, as we shall demonstrate.

As a side note, each of the assumptions above can be relaxed independently. The first approximation (i) can be amended within the linear formalism by means of the matrix method \citep[e.g.][]{Kalnajs1977Dynamics,Weinberg1989SelfGravitating,Chavanis2012Kinetic,Heyvaerts2017Dressed}. The second and third approximations (ii, iii) can also be relaxed within the linear formalism, resulting in a general time-dependent torque formula \citep{Weinberg2004Timedependent,Banik2021SelfConsistent}, which can also incorporate collective effects \citep[e.g.][]{Murali1999Transmission,Rozier2022Constraining,Dootson2022}. The last linear approximation (iv) causes an issue near the resonances, which can be worked around instead by the method of averaging \citep{lichtenberg1992regular}. Dynamical friction in the nonlinear regime was studied by \cite{Tremaine1984Dynamical} in the time-asymptotic limit and by \cite{Chiba2022Oscillating,Banik2022Nonperturbative} in the slow limit. Both limits were removed in \cite{Chiba2023GeneralFastSlowRegime}. \cite{Hamilton2022BarResonanceWithDiffusion} included the effect of diffusion and demonstrated the recovery of the linear regime in the strongly diffusive limit.

The LBK formula (\ref{eq:LBK}) states that dynamical friction arises from the gradient in the distribution function at the resonances: $\vn \cdot \vOmega - \nphi \Omegap = 0$. The physical picture of this is analogous to Landau damping \citep{Landau1946}: near a resonance, particles initially on the lower energy side of the resonance will, on average, gain energy from the perturber, while those on the higher energy side of the resonance will give up energy to it. Dynamical friction occurs when the former outnumbers the latter.

More specifically, the LBK torque (\ref{eq:LBK}) depends on the slope of the distribution function in action space $\vJ$ in the direction parallel to the resonant vector $\vn$. The reason for this is best understood via the fast-slow angle-action variables \citep[e.g.][]{LyndenBell1979BarMechanism,Tremaine1984Dynamical}. In the frame rotating at $\Omegap$, unperturbed orbits precess with the resonant frequency $\Omegas \equiv \vn \cdot \vOmega - \nphi \Omegap$ \citep[see the illustration by][]{Weinberg2007BarHaloInteraction}. As the resonance is approached, this precession frequency becomes much smaller than the individual orbital frequencies $\vOmegaf \equiv (\Omegar,\Omegapsi)$, so the dynamics splits into slow and fast components. The orbital phase of the respective motions is given by the slow angle $\thetas \equiv \vn \cdot \vtheta - \nphi \int \drm t ~\Omegap$ and the fast angles $\vthetaf \equiv (\thetar,\thetapsi)$. Making a canonical transformation to these new angles, one finds that their conjugate slow-fast actions are $(\Js, \Jfo,\Jft) \equiv (\Lz / \nphi, \Jr-\nr\Lz/\nphi, ~L-\npsi\Lz/\nphi)$. Because actions are adiabatic invariants, when averaged over time longer than the periods of the fast motions, $\vOmegaf^{-1}$, only the slow action may change, while the fast actions are effectively constant. It is easily verified that these fast actions constitute vectors perpendicular to $\vn$,
\begin{align}
  \vJ \times \vn = (\nphi\Jfo, ~\nphi\Jft, ~\npsi\Jfo-\nr\Jft).
  \label{eq:J_perpendicular}
\end{align}
Since fast actions are conserved, it follows that secular evolution in action space occurs only in the direction parallel to $\vn$. The factor $\vn \!\cdot\! \pd f/\pd\vJ$ in the LBK formula represents the fact that dynamical friction is caused by this slow rectilinear evolution of the particle's distribution along the resonant vectors $\vn$ \citep[see related discussion by ][]{Fouvry2015Application}.

\section{Results}
\label{sec:results}

Using the spinning dark halo introduced in Section \ref{sec:model_spinning_halo}, we now compute the torque on galactic bars and investigate its dependence on the halo's rotation. The potential perturbation by the bar is modeled by a simple quadrupole rotating at $\Omegap=36 \Gyr^{-1}$ \citep[for details, see][]{Chiba2022Oscillating}. In this model, the Fourier coefficients $\hPhin$ are non-zero only for $\nphi=2$ and $\npsi=2,0,-2$. The possible values of $\nr$ in our standard Hernquist potential are $\nr \geq -1$ for $\npsi=2$, $\nr \geq 1$ for $\npsi=0$, and $\nr \geq 2$ for $\npsi=-2$ \citep[][]{Chiba2022Oscillating}.

\subsection{LBK torque by spinning dark halos}
\label{sec:LBK_spinning_halos}

The LBK torque (\ref{eq:LBK}) by a spinning halo (\ref{eq:DF_halo}) is 
\begin{align}
  \tau_{\rm LBK} =& \left(2 \pi \right)^3 \sum_\vn \nphi \!\int\!\drm^3 \vJ 
  \vn \!\cdot\! \left[\left(1 + g\right) \frac{\pd f_{+}}{\pd \vJ} + f_{+} \frac{\pd g}{\pd \vJ}\right] \nonumber \\
  &\times |\hPhin(\vJ)|^2 \pi \delta\!\left(\vn \cdot \vOmega - \nphi \Omegap\right).
  \label{eq:LBK_spin}
\end{align}
The first term in the square bracket describes the gradient in the original non-rotating DF $f_{+}$ weighted by $(1+g)$, i.e. the redistribution of mass in $\Lz$. We denote the torque arising from this term as $\tau_{f_{+}}$. The second term describes the gradient in the DF due to the reweighting itself. We refer to the torque resulting from this term as $\tau_{g}$. To clarify the origin of dynamical friction, we present the two torques separately. Note that the LBK torque is linear in $\Lambda$, so in practice one only needs to compute the torque at $\Lambda=0$ and its derivative with respect to $\Lambda$.

\begin{figure}
  \begin{center}
    \includegraphics[width=8.5cm]{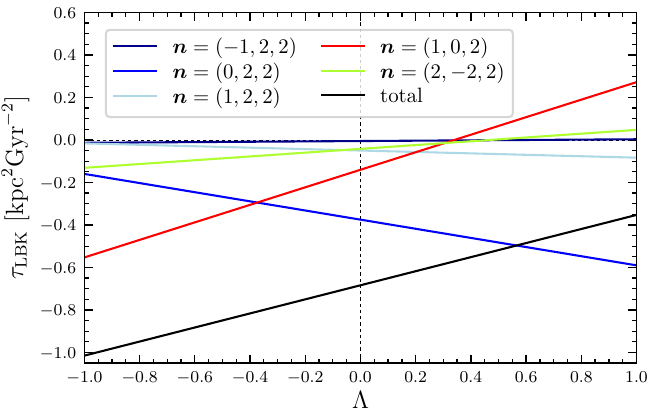}
    \caption{LBK torque as a function of halo spin $\Lambda$ in a Hernquist halo. The total torque (black) is the sum over all resonances $\npsi=2,0,-2$ up to $\nr = 10$.}
    \label{fig:LBK_spin_total}
  \end{center}
\end{figure}

Fig.~\ref{fig:Phin_LBK} plots the LBK torque as a function of the halo's spin $\Lambda$ (bottom) along with the amplitude of the perturbation $|\hPhin|$ (top) for three different resonances. The left column plots the corotation resonance $\vn=(0,2,2)$. This resonance represents the series of resonances with $\npsi=2$, including the Lindblad resonances $\vn=(\pm1,2,2)$. As shown in the top panel, the amplitude of these resonances increases monotonically with $\Lz$.\footnote{It is important to note that, in spherical systems, the orbital frequencies $(\Omegar,\Omegapsi)$, and thus the resonance condition, only depend on $\Jr$ and $L$, but not $\Lz$, which determines the orbital inclination $\beta$. Hence, strange as it may sound, retrograde orbits ($\beta>\pi/2$) may well be trapped into the ``corotation'' resonance of the bar ($\Omegapsi=\Omegap$), even though they are not corotating in the cylindrical azimuthal angle! The volume of phase space trapped in the corotation resonance declines monotonically towards large $\beta$ but only vanishes at $\beta=\pi$ (Fig.~\ref{fig:Phin_LBK}). [The general dependence of each resonance on $\beta$ is given by the Wigner's rotation matrix \citep{Tremaine1984Dynamical}. See also \cite{Breen2021kinematicII} Appendix B for a pedagogical explanation.]} Hence, increasing the prograde population at the expense of retrograde orbits enhances the net dynamical friction. This is represented by the decrease in $\tau_{f_{+}}$ with the halo's spin $\Lambda$. (Note that dynamical friction is a negative torque, so a decrease in the torque means an increase in friction.) In contrast, $\tau_{g}$ increases with $\Lambda$: spinning the halo forward results in a positive gradient with respect to $\Lz$ near $\Lz=0$, which in turn generates a positive torque. Since the change in $\tau_{f_{+}}$ outweighs the change in $\tau_{g}$, the total dynamical friction increases with $\Lambda$\footnote{This holds for all $\npsi=2$ resonances but the inner Lindblad resonance, at which the change in $\tau_g$ surpass the change in $\tau_{f_{+}}$ (see Fig.~\ref{fig:LBK_spin_total}).}. This conclusion agrees with \cite{weinberg1985evolution}.

For other resonances, however, the consequences are different. The middle column shows the case for the direct radial resonance $\vn=(1,0,2)$, where an orbit completes two radial oscillations during a bar period, $\Omegar-2\Omegap=0$ \citep{Weinberg2007BarHaloInteraction}. This resonance represents the series of resonances with $\npsi=0$, and is not to be confused with the outer Lindblad resonance, which occurs at $\Omegar+2(\Omegapsi-\Omegap)=0$. The amplitude of the direct radial resonance peaks at $\Lz=0$ and decays symmetrically towards $\Lz=\pm L$ where it vanishes. Hence, this resonance is most relevant for orbits perpendicular to the galactic plane but is irrelevant for in-plane disk orbits. Due to its symmetric amplitude about $\Lz=0$, transferring mass between retrograde and prograde phase-space has no effect on $\tau_{f_{+}}$: prograde and retrograde orbits resonate with the perturber at equal strength, so varying their mass ratio does not affect the net friction. However, the gradient at $\Lz=0$ caused by the rotation generates a large torque because the resonance is strongest there. Thus, $\tau_g$ rises sharply with $\Lambda$. The net torque from this resonance increases with $\Lambda$ and even becomes positive beyond $\Lambda \sim 0.35$.

Finally, the right column plots the torque by $\vn=(2,-2,2)$, which is the strongest resonance among those with $\npsi=-2$. These resonances are strongest when the orbit is counter-rotating with respect to the bar (i.e. the amplitude shows an increase towards negative $\Lz$). Hence, both $\tau_{f_{+}}$ and $\tau_g$ increase with $\Lambda$.

Fig.~\ref{fig:LBK_spin_total} shows the torque by the main resonances and the total torque, including all resonances up to $\nr = 10$. The torques by the corotation resonance $\vn=(0,2,2)$, the outer Lindblad resonance $\vn=(1,2,2)$, and other higher order resonances $\nr \geq 2$ with $\npsi=2$ not presented in Fig.~\ref{fig:LBK_spin_total} decrease with the halos' spin $\Lambda$. Meanwhile, the torques by the inner Lindblad resonance $\vn=(-1,2,2)$\footnote{In a potential with a Hernquist halo alone, the ILR of the current Galactic bar would lie at a very small angular momentum ($\sim 10 \kpc^2\Gyr^{-1}$), so its contribution to dynamical friction is small. The ILR becomes significant in a model with a much higher concentration of mass at the inner region (see Appendix \ref{sec:signular_isothermal_sphere}).}, the direct radial resonance $\vn=(1,0,2)$\footnote{This resonance also makes a significant contribution to the bar instability, comparable to that of the corotation resonance \citep{Breen2021kinematicII}.}, the retrograde resonance $\vn=(2,-2,2)$, and all other resonances with $\npsi=0,-2$ increase with $\Lambda$. The impact of the latter exceeds the former, with the result that the total torque increases (dynamical friction weakens) with $\Lambda$. The LBK formula thus correctly accounts for the reduced dynamical friction on bars in spinning halos found in $N$-body simulations at the secular phase \citep[][]{Long2014Secular}.

\subsection{Variation with steepness parameter}
\label{sec:smoothing_DF}

So far, we have set the halo rotating using equation (\ref{eq:DF_g_tanh}) with the steepness parameter set to $\chi=3$. We now check that our qualitative conclusion is independent of our choice of $\chi$.

\begin{figure}
  \begin{center}
    \includegraphics[width=8.5cm]{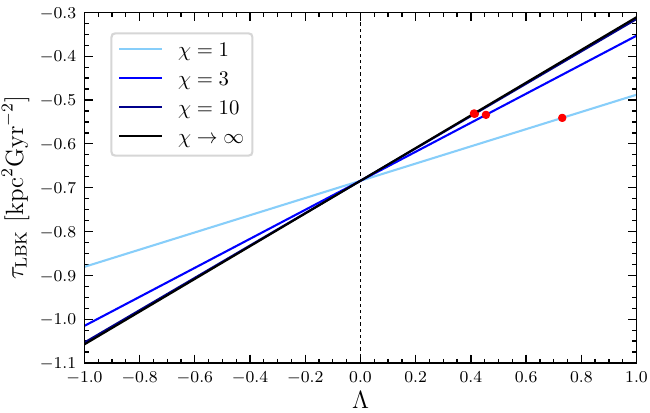}
    \caption{The dependence of dynamical friction on the halo's spin $\Lambda$ for various steepness $\chi$ of the rotating distribution function (equation \ref{eq:DF_g_tanh}). The red dots mark the values of $\Lambda$ for each $\chi$ that yield $\lambda=0.1$.}
    \label{fig:LBK_chi}
  \end{center}
\end{figure}

Fig.~\ref{fig:LBK_chi} plots the LBK torque as a function of the halo spin $\Lambda$ for a variety of $\chi$, where $\chi \rightarrow \infty$ corresponds to a discontinuous DF (equation \ref{eq:DF_g_sgn}). A reduction in dynamical friction with $\Lambda$ is observed irrespective of $\chi$, suggesting the $\chi$ is qualitatively unimportant. Our result also clarifies that the behavior observed in past simulations using $\chi \rightarrow \infty$ \citep[e.g.][]{Long2014Secular} is not an artifact of the discontinuity in the DF.

Quantitatively, a larger $\chi$ causes a larger impact on dynamical friction. Note however that a larger $\chi$ also implies a larger total angular momentum $\lambda$ (cf. Fig.~\ref{fig:chi_lambda}). To give a fair comparison, we marked with red dots the torque expected from halos with equal total angular momentum $\lambda = 0.1$: the larger the $\chi$, the smaller the $\Lambda$ must be in order to keep $\lambda$ constant. We find that, despite the same $\lambda$, the impact of spinning the halo becomes more prominent for a steeper distribution (larger $\chi$). Hence, simulations employing a discontinuous DF are likely overestimating the impact of the halo's spin, though the effect is small.

\begin{figure}
  \begin{center}
    \includegraphics[width=8.5cm]{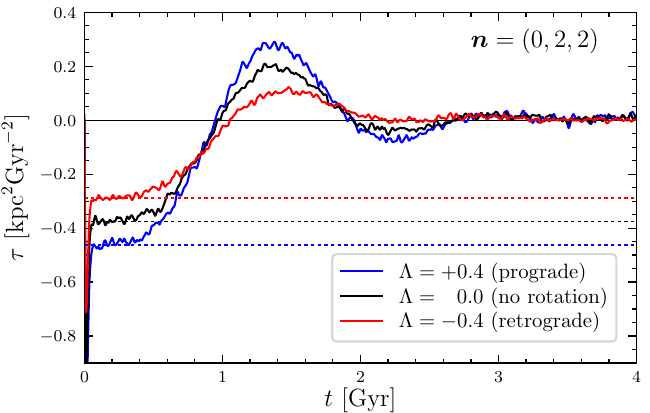}
    \includegraphics[width=8.5cm]{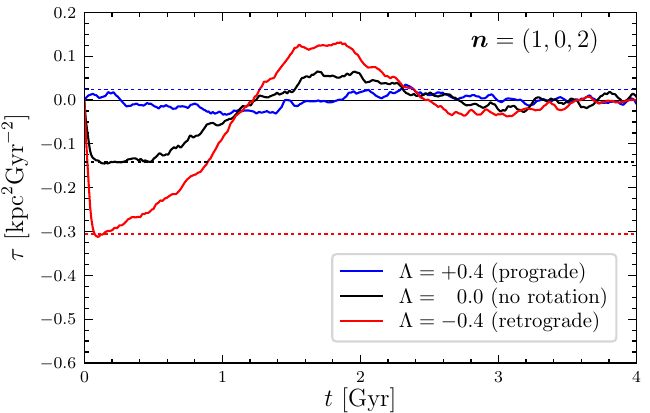}
    \includegraphics[width=8.5cm]{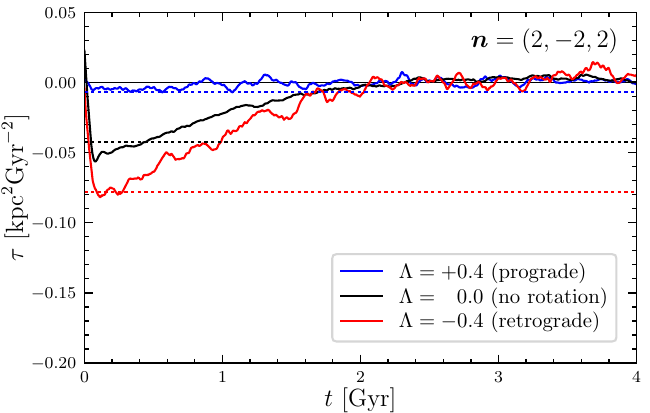}
    \caption{Torque on the bar from individual resonances computed numerically using the angle-averaged Hamiltonian. The dotted lines represent the torque predicted by the linear LBK formula (Fig.~\ref{fig:Phin_LBK}). Note the opposite dependence of the torque on $\Lambda$ for the corotation resonance (top) and the other two resonances (middle and bottom).}
    \label{fig:LBK_vs_nonlinear}
  \end{center}
\end{figure}

\subsection{Comparison with test-particle simulations}
\label{sec:test_particle_simulation}

To verify our linear calculations, we compute the full nonlinear torque from each individual resonance using the method described in \cite{Weinberg2007BarHaloInteraction} and \cite{Chiba2023GeneralFastSlowRegime}. In short, the contribution to the torque from each resonance is computed by integrating test particles in the slow angle-action space using the angle-averaged Hamiltonian, which only retains a single resonant term of the perturbation.

Fig.~\ref{fig:LBK_vs_nonlinear} shows the torque from individual resonances as a function of time for three halo spins. The bar is introduced instantly at $t=0$ and rotates at a fixed pattern speed. The simulations (solid curves) agree with the linear LBK formula (dotted lines) for approximately half a libration period ($\sim 500 \Myr$), after which the evolution becomes nonlinear. The simulations reconfirm that a forward-spinning halo ($\Lambda>0$, blue) strengthens dynamical friction by the $\vn=(0,2,2)$ resonance but weakens friction by the $\vn=(1,0,2)$ and $\vn=(2,-2,2)$ resonances.

The nonlinear torque decays at late times because the resonant orbits slowly phase mix, thereby erasing the gradient in the DF near the resonances \citep{Chiba2022Oscillating,Banik2022Nonperturbative}. In a self-consistent simulation (Section \ref{sec:N_body_simulation}), the torque remains non-zero because (i) the resonances continuously move \citep{Chiba2023GeneralFastSlowRegime}, and (ii) the original gradient in the DF can be recovered through orbital diffusion \citep{Hamilton2022BarResonanceWithDiffusion}.

\begin{figure}
  \begin{center}
    \includegraphics[width=8.5cm]{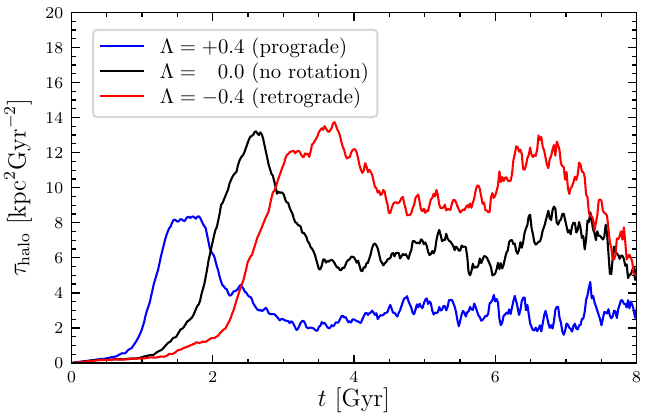}
    \caption{Torque on the dark halo measured in $N$-body simulations of a barred galaxy.}
    \label{fig:nbody_dLzdt}
  \end{center}
\end{figure}

\subsection{Comparison with $N$-body simulations}
\label{sec:N_body_simulation}

We also compare qualitatively the prediction by the LBK theory with a set of $N$-body simulations of a galaxy consisting of a stellar disk and a live dark matter halo. The details of the setup can be found in \cite{Kataria2022Effects}. Since the impact of the halo's spin is only weakly dependent on the steepness parameter $\chi$ (Section \ref{sec:smoothing_DF}), here we adopt the discontinuous DF ($\chi \rightarrow \infty$): we set the halo rotating forward (backward) by reversing the velocities of a certain fraction $\Lambda$ of retrograde (prograde) orbits. The impact of the discontinuity on the formation and evolution of bars is explored by Kataria \& Shen (2023, in prep.).

Fig.~\ref{fig:nbody_dLzdt} shows the torque on the halo, which is a proxy for the dynamical friction on the bar. During the late secular phase, the torque is stronger in retrograde halos ($\Lambda < 0$) in agreement with the LBK formula. At early times, however, the torque is stronger in prograde halos ($\Lambda > 0$) because the bar grows more rapidly \citep{Kuijken1994Lowered,Saha2013Spinning,Long2014Secular}. This implies that the rotation of the halo not only affects the secular torque but also the bar instability, as suggested by linear response theory \citep[e.g.][]{Kalnajs1977Dynamics,Weinberg1991Instability,Pichon1997Numerical}. \cite{Rozier2019Mapping}, in particular, analyzed the stability of rotating Plummer spheres and identified unstable $m=2$ modes. We conjecture that our rotating dark halo was similarly unstable to $m=2$ modes and hence formed a `dark bar' \citep{Athanassoula2007BarInHalo,Petersen2016DarkMatterTrapping,Collier2019DarkMatterBars}, which subsequently fostered (or retarded) the formation of the stellar bar.

\subsection{Observational implication}
\label{sec:observational_implication}

Finally, we briefly discuss the observational implication of the dark halo's spin. Although the halo's spin cannot be observed directly, its secular effect on bar evolution can be inferred from the current properties of barred galaxies \citep[e.g.][]{Ansar2023Modelling}. In particular, both simulations \citep{Mayer2004LSBG} and observations \citep{Honey2016Nearinfrared,CervantesSodi2017Do} find that bars in low surface brightness (LSB) galaxies tend to be short, implying that they experienced weak dynamical friction since friction makes the bar slow and simultaneously grow in length \citep[e.g.][]{Martinez2006Evolution}. These LSB galaxies, on the other hand, are known to form in dark halos with high spin \citep{Kim2013How,DiCintio2019LSBGinNIHAO,PerezMontano2022formation} because the spin of dark matter and baryons are correlated \citep{Chen2003Angular,Sharma2005Angular,Obreja2022first}\footnote{See however \cite{Jiang2019DMspinGalaxyspin} for a negative evidence.}: the higher the halo spin, the larger the angular momentum/scale radius of the stellar disk, and hence the lower its surface density \citep{Fall1980Formation,Dalcanton1997Formation,Jimenez1998Galaxy}. Hence, the short immature bars observed in LSB galaxies could be due, in part, to the large spin of the halo in line with our theoretical prediction.

\section{Conclusion}
\label{sec:conclusion}

This paper revisited the issue of the contradicting predictions made in the past by linear theory (LBK formula) and $N$-body simulations regarding the impact of the dark halo's rotation on dynamical friction. We reevaluated the LBK formula and demonstrated that, contrary to past studies, it predicts a reduced dynamical friction by halos with net prograde rotation, in agreement with $N$-body simulations.

The LBK formula states that dynamical friction on the perturber is caused by the gradient in the particle's distribution function near the resonances. Spinning the dark halo forward modifies the gradient of the distribution function in two ways: Firstly, it enhances the density gradient in the prograde phase space by simply increasing the overall prograde population. The gradient in the retrograde phase space is conversely reduced. This causes a net increase in dynamical friction, since the main resonances (e.g. corotation resonance) have larger amplitude in the prograde phase space (see $\tau_{f_{+}}$ in Fig.~\ref{fig:Phin_LBK}). Secondly and more importantly, spinning the halo results in a positive gradient in the distribution function with respect to $\Lz$, typically near the origin $\Lz=0$. This positive gradient generates a positive torque on the perturber from all resonances, thus weakening dynamical friction (see $\tau_g$ in Fig.~\ref{fig:Phin_LBK}). The net change in dynamical friction is determined by the sum of these two opposing effects.

In contrary to the seminal work by \cite{weinberg1985evolution}, we found that the latter effect outweighs the former with the result that the overall dynamical friction weakens with the halo's spin. A key role is played by the direct radial resonances ($\npsi=0$), where the radial frequency of orbits directly couples with the pattern speed \citep{Weinberg2007BarHaloInteraction}. These resonances are highly responsive against the halo's rotation since they are strongest at $\Lz=0$, which is where the halo's distribution typically bears the largest gradient in $\Lz$. We found that the total reduction of dynamical friction by these resonances, together with the $\npsi=-2$ resonances and the inner Lindblad resonance, exceeds the amplification of dynamical friction by the rest of the $\npsi=2$ resonances, including the corotation resonance and the outer Lindblad resonance. We verified our linear calculation with the nonlinear torque computed using the angle-averaged Hamiltonian (Section \ref{sec:test_particle_simulation}), and confirmed that the conclusion holds for halos with a variety of density profiles (Appendix \ref{sec:signular_isothermal_sphere}) and $\Lz$ distribution (Section \ref{sec:smoothing_DF}, Appendix \ref{sec:position_of_gradient}).

In this paper, we have restricted our analysis to spinning halos constructed based on spherical isotropic models. However, simulations by others have found similar effect of the halo's spin even with nonspherical \citep{Collier2018SpinningHalo} or anisotropic halos \citep{debattista2000constraints}. It would nevertheless be of interest to investigate whether and how deviation from sphericity or isotropy might change the relative contributions from each resonance.

An important task in the future will be to provide constraints on the spin of the dark halo in the Milky Way, which is required to calculate the event rate in direct detection experiments \citep[e.g.][]{Green2001Weakly}. \cite{Obreja2022first} recently estimated the spin of the Galactic dark halo from its correlation with that of the stellar components, in particular the stellar halo. We propose another possibility, utilizing the recent measurement of the bar's slowing rate \citep{Chiba2020ResonanceSweeping}. A natural approach would be to quantify the impact of the halo's spin on the bar's slowing rate using a self-consistent dynamical model of our Galaxy fitted to the observational data \citep[][]{Cole2017centrally,Binney2023Selfconsistent}. Beyond modeling dynamical friction, this will require estimates on (i) the bar's moment of inertia, which we may infer from $N$-body simulations \citep[][]{Beane2022StellarBars}, and (ii) the rate of angular momentum exchange between the bar and other baryons, in particular the gas in the disk. Gas typically yields angular momentum to the bar \citep{VillaVargas2010Gas,Athanassoula2013BarGas,Beane2022StellarBars} and consequently sinks towards the galactic center, where it forms the nuclear stellar disk \citep[NSD,][]{Launhardt02,Matsunaga2015CepheidsNSD,Schoenrich15}. The torque from the gas can be estimated from the gas infall rate \citep{Sormani2019MassInflowRate}, or from the mass of the NSD \citep{Sormani2022NSD} together with its age \citep{Sanders2023epoch}, or possibly from the chemical history of the NSD \citep{Friske2023Chemical}.

\section*{Acknowledgements}

We thank M.Weinberg for fruitful discussions and for confirming our result. We are also grateful to S.Rozier and the referee C.Pichon for useful comments. R.C. thanks support by the the Natural Sciences and Engineering Research Council of Canada (NSERC), [funding reference \#DIS-2022-568580]. S.K.K. thanks the support by the National Key R\&D Program of China under grant No. 2018YFA0404501; by the National Natural Science Foundation of China under grant Nos. 12025302, 11773052, 11761131016; by the ``111'' Project of the Ministry of Education of China under grant No. B20019; and by the China Manned Space Project under grant No. CMS-CSST-2021-B03.

\section*{Data availability}

The codes used to produce the results are available from the corresponding author upon request.



\bibliographystyle{mnras}
\bibliography{./references}



\appendix

\section{Distribution of angular momentum in the Galactic meridian plane}
\label{sec:spatial_distriution_of_AM}

\begin{figure}
  \begin{center}
    \includegraphics[width=8.2cm]{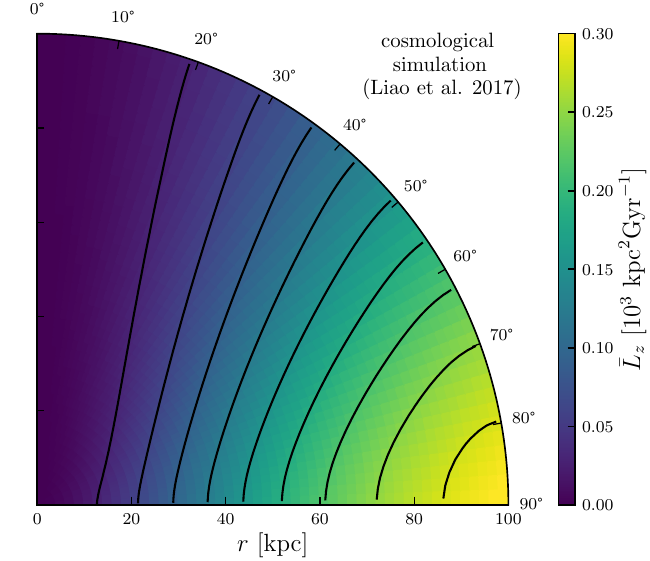}
    \includegraphics[width=8.2cm]{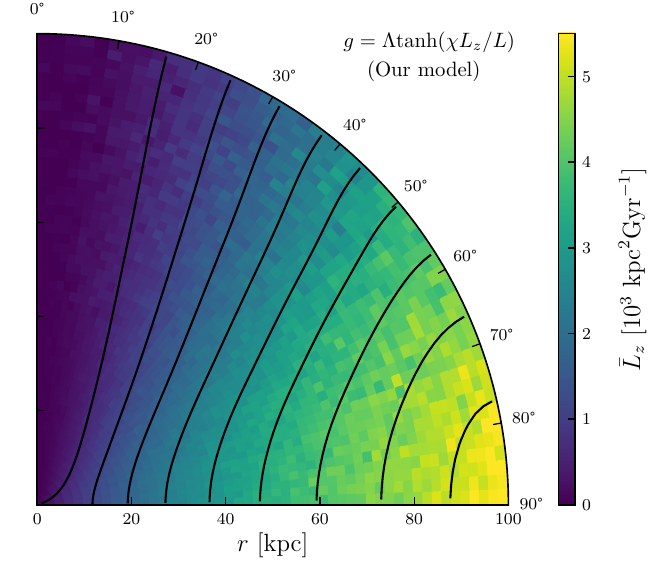}
    \includegraphics[width=8.2cm]{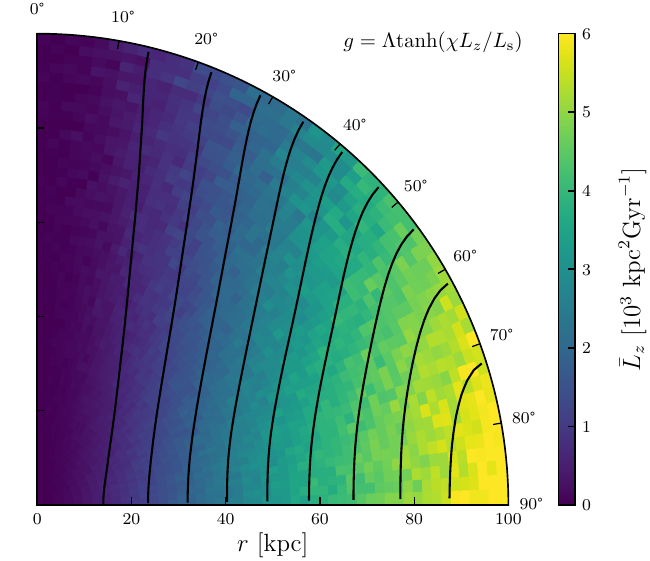}
    \caption{Mean angular momentum of spinning dark halos in the meridian plane ($r,\vartheta$). Top panel: Empirical model by Liao et al. (2017) based on cosmological simulation. Middle panel: Our model with a spin function $g$ depending on the ratio $\Lz/L$ (equation \ref{eq:DF_g_tanh}). Bottom panel: Model with $g$ depending only on $\Lz$ (equation \ref{eq:DF_g_tanh_LzLs}).}
    \label{fig:Lz_rth}
  \end{center}
\end{figure}

In Section \ref{sec:model_spinning_halo}, we compared the two models of spinning dark halos (equations \ref{eq:DF_g_tanh} and \ref{eq:DF_g_tanh_LzLs}) by observing the angular momentum averaged over spherical shells as a function of radius. Here, we take the comparison a step further and look into the angular momentum averaged over azimuthal rings as a function of radius $r$ and polar angle $\vartheta$.

\cite{Liao2017Universal} analyzed the dark halos in the Bolshoi cosmological simulation \citep{Klypin2011Bolshoi} and found that the distribution of angular momentum in the $(r,\vartheta)$ plane can be fitted with the following empirical function
\begin{align}
  \bar{L}_z(r,\vartheta) = j_{\rm s} \frac{(\tilde{r}/\rs)^2}{(1+\tilde{r}/\rs)^4} \sin^2(\vartheta/\vartheta_{\rm s}),
  \label{eq:Liao_model}
\end{align}
where $\tilde{r} \equiv r / r_{\rm vir}$, and $j_{\rm s}$, $\rs$, and $\vartheta_{\rm s}$ are free parameters. They fit this function to multiple dark halos and derived the scaling relation between the free parameters and the virial mass of the halo.

Fig.~\ref{fig:Lz_rth} top panel shows the distribution of $\bar{L}_z$ from their empirical model using the virial mass of our Galaxy ($M_{\rm vir} = 1.5 \times 10^{12} \Msun$). The distribution is skewed towards the equatorial plane, which is naturally expected as $\Lz$ is largest when the angular-momentum vector is aligned with the $z$ axis. 

The middle panel shows $\bar{L}_z$ of our spinning halo model (equation \ref{eq:DF_g_tanh}) obtained by Monte Carlo integration. The distribution is surprisingly similar to the empirical model proposed by \cite{Liao2017Universal}. The absolute value of $\bar{L}_z$ is much higher than their model since we set $\Lambda=1$. A quantitative match would require $\Lambda \sim 0.06$. 

The bottom panel shows $\bar{L}_z$ of the spinning halo with a universal scale for the spin function $g$ (equation \ref{eq:DF_g_tanh_LzLs}). Compared to our model as well as that of \cite{Liao2017Universal}, the distribution depends more on the radial distance from the $z$ axis (i.e. $R=r\sin\vartheta$) rather than the polar angle $\vartheta$. This reflects the fact that the spin function $g$ in this model depends only on $\Lz$, which sets the range of $R$ an orbit can explore: the smaller the $\Lz$, the smaller the minimum $R$ an orbit may reach. In contrast, $g$ in our model is a function of the orbital inclination $\beta$, which determines the orbital range in $\vartheta$.

\section{Singular isothermal sphere}
\label{sec:signular_isothermal_sphere}

\begin{figure}
  \begin{center}
    \includegraphics[width=8.5cm]{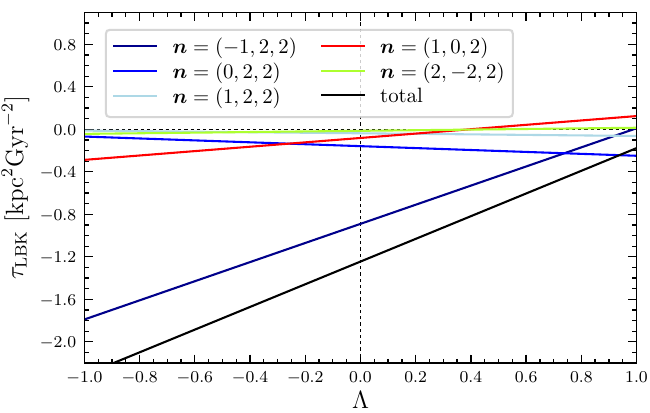}
    \caption{LBK torque as a function of the halo's spin $\Lambda$ in a singular isothermal sphere.}
    \label{fig:LBK_spin_total_isothermal}
  \end{center}
\end{figure}

In this section, we repeat the calculation of the LBK formula using the singular isothermal sphere, which was employed in \cite{weinberg1985evolution}. The singular isothermal sphere has the following density and potential:
\begin{align}
  \rho(r) = \frac{\Ms}{4\pi\rs r^2}, ~~~~
  \Phi(r) = \frac{G\Ms}{\rs} \ln \left( \dfrac{r}{\rs} \right),
  \label{eq:halo_isothermal}
\end{align}
where $\Ms$ is the mass enclosed within $\rs$. The isothermal sphere has a flat rotation curve $\vc = \sqrt{G\Ms/\rs}$ and a uniform velocity dispersion $\sigma=\vc/\sqrt{2}$, hence the name. The circular speed $\vc$ is the single independent parameter of the model, and we set this to $235 \kms$. The distribution function is $f(E) \propto \exp(- E/\sigma^2)$. Since the isothermal sphere has an infinite mass, we truncate the distribution at $E_{\rm max} = \Phi(r_{\rm max})$ where $r_{\rm max}=10^3 \kpc$.

Fig.~\ref{fig:LBK_spin_total_isothermal} plots the torque on the bar by the singular isothermal sphere. Similar to the Hernquist halo, the total dynamical friction weakens with the halo's spin $\Lambda$, suggesting that the discrepancy between our calculation and that of \cite{weinberg1985evolution} is not due to the difference in the halo's density profile. The relative contribution from each resonance is, however, different from the Hernquist model. In particular, the inner Lindblad resonance [ILR, $\vn=(-1,2,2)$] now dominates the torque. This is because the isothermal sphere has a higher concentration of mass in the inner region, which raises the orbital frequencies there and thus allows the ILR to expand and cover a larger phase space. We caution though that, while the isothermal sphere places the ILR at a realistic position, the mass of dark matter in the inner region near the ILR is significantly overestimated. A proper quantification of the relative importance of each resonance will require a model for the distribution function of the dark halo in equilibrium under the combined potential of the dark halo and the stellar disk.

We also explored other double power-law models $\rho \propto r^{-\gamma} (r+\rs)^{\gamma-\eta}$, in particular, the truncated \citetalias{Navarro1997Universal} model $(\gamma=1,\eta=3)$ and the \cite{Jaffe1983} model $(\gamma=2,\eta=4)$ but found similar results. The former yielded results closer to those of the Hernquist model $(\gamma=1,\eta=4)$, while the latter was closer to the singular isothermal sphere $(\gamma=2,\eta=2)$, indicating that dynamical friction depends mainly on the density slope at the inner region $\gamma$, where the low-order resonances reside.

\section{Varying the location of discontinuity}
\label{sec:position_of_gradient}

\begin{figure}
  \begin{center}
    \includegraphics[width=8.5cm]{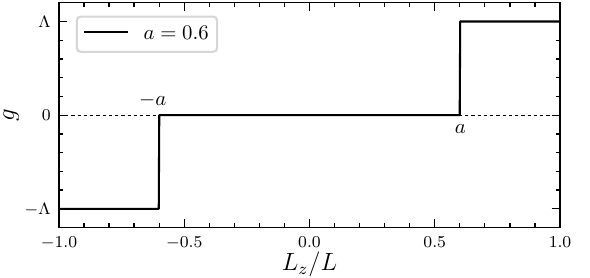}
    \caption{The spin function $g$ (equation \ref{eq:DF_g_W}) corresponding to the Weinberg's rotating halo model.}
    \label{fig:g_Lz_W}
  \end{center}
\end{figure}

In this paper, we adopted a spinning halo model in which the DF bears the largest gradient in $\Lz$ at $\Lz=0$ (Section \ref{sec:LBK_spinning_halos}). In this section, we check whether our conclusion holds also for spinning models with large gradients (in particular, discontinuities) in $\Lz$ away from the origin. Such a model was in fact used by \cite{weinberg1985evolution}, who set the halo rotating by reversing retrograde orbits with $\pi/2(2-x) \leq \beta \leq \pi$ where $0 \leq x \leq 1$ and $\beta=\cos^{-1}(\Lz/L)$ is the orbital inclination. Taking the cosine, the range transforms to $-1 \leq \Lz/L \leq -a$ where $a \equiv -\cos\left[\pi/2(2-x)\right] \geq 0$. Reversing orbits in this range is equivalent to setting the odd DF $g$ (equation \ref{eq:DF_halo_odd}) as
\begin{align}
  g(L,\Lz) = \Lambda \left[\Theta(\Lz/L-a) + \Theta(\Lz/L+a) - 1 \right],
  \label{eq:DF_g_W}
\end{align}
where $\Theta$ is the Heaviside step function. We plot this function in Fig.~\ref{fig:g_Lz_W}. The function has two discontinuities at $\Lz/L =\pm a$. In the limit $a=0$, the two discontinuities merge, and the function reduces to equation (\ref{eq:DF_g_sgn}).

\begin{figure}
  \begin{center}
    \includegraphics[width=8.5cm]{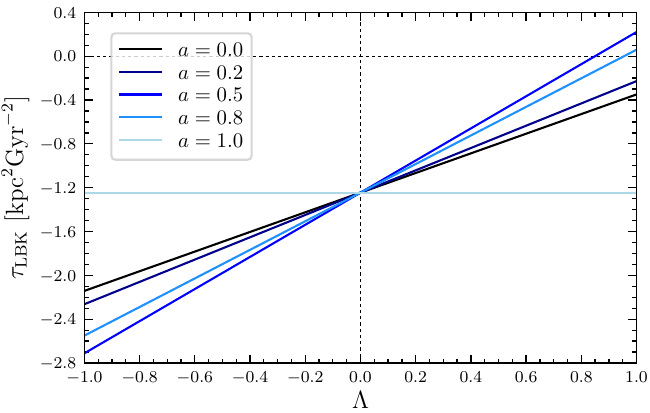}
    \caption{The LBK torque by an isothermal sphere with two discontinuities away from the origin (equation \ref{eq:DF_g_W}).}
    \label{fig:LBK_a_isothermal}
  \end{center}
\end{figure}

Fig.~\ref{fig:LBK_a_isothermal} shows the LBK torque by the isothermal sphere set rotating using equation (\ref{eq:DF_g_W}). We find that, regardless of $a$, dynamical friction on the bar decreases with $\Lambda$. Interestingly, the impact of the halo's spin first increases with $a$. This is because increasing $a$ shifts the two discontinuities towards $\beta=0$ and $\pi$ respectively, so the positive torque due to the discontinuities (i.e. $\tau_g$) by the $\npsi=\pm2$ resonances increases. Beyond $a\sim0.5$, however, the slope begins to flatten because the range of orbits reversed in motion becomes narrower. The impact of the halo's spin vanishes at $a=1$ where there is no net rotation.


\bsp	
\label{lastpage}
\end{document}